# Development and optimization of large-scale integration of 2D material in memristors.

Clotilde Ligaud[1], Lucie Le Van-Jodin[1], Bruno Reig[1], Pierre Trousset[1], Paul Brunet[1], Michaël Bertucchi[1], Clémence Hellion[1], Nicolas Gauthier[1], Le Van-Hoan[1], Hanako Okuno[2], Djordje Dosenovic[2], Stéphane Cadot[1], Remy Gassilloud[1], Matthieu Jamet[3]

[1] Univ. Grenoble Alpes, CEA, Leti, F-38000 Grenoble, France
[2] Univ. Grenoble Alpes, CEA, IRIG, MEM, F-38000 Grenoble, France
[3] Univ. Grenoble Alpes, CEA, CNRS, Grenoble INP, IRIG, Spintec, F-38000 Grenoble, France

E-mail: lucie.levan-jodin@cea.fr

**Abstract**

Two-dimensional (2D) materials like transition metal dichalcogenides (TMD) have proved to be serious candidates to replace silicon in several technologies with enhanced performances. In this respect, the two remaining challenges are the wafer scale growth of TMDs and their integration into operational devices using clean room compatible processes. In this work, two different CMOS-compatible protocols are developed for the fabrication of $MoS_2$-based memristors, and the resulting performances are compared. The quality of $MoS_2$ at each stage of the process is characterized by Raman spectroscopy and x-ray photoemission spectroscopy. In the first protocol, the structure of $MoS_2$ is preserved during transfer and patterning processes. However, a polymer layer with a minimum thickness of 3 nm remains at the surface of $MoS_2$ limiting the electrical switching performances. In the second protocol, the contamination layer is completely removed resulting in improved electrical switching performances and reproducibility. Based on physico-chemical and electrical results, the switching mechanism is discussed in terms of conduction through grain boundaries.

## 1. Introduction

Since the discovery of the unique properties of graphene [1], 2D materials have been holding great promises for the development of high-performance electronic devices. In particular, the intrinsic properties of $MoS_2$ could allow the development of flexible devices [2], sensors [3], [4], high responsivity ($10^4$ mA/W) phototransistors [5], non-volatile memory devices [6], high cut-off frequency RF switches [[7], [8], [9], [10]] or transistors [11], [12].

For memory devices, $MoS_2$ has been integrated in nanoscale vertical metal-insulator-metal (MIM) structure resulting in high performances devices [13] with low switching voltage, fast switching speed and high ON/OFF current ratio enabling the manufacture of multi-level devices [14], [15] and opening the way to the development of RF switches [13], [16].

However, despite recent promising results, the atomic mechanisms for the formation of ON and OFF states in these devices are not yet well understood in particular because of a lack of systematic and reproducible data. Several hypotheses have been put forward. Wang et al. [17] explain the switching behaviour in partially oxidised $MoS_2$ by the formation of a conductive filament made of sulphur vacancies which is formed or resorbed by the diffusion of oxygen ions. In the devices presented by Yin et al. [18], $MoS_2$ was inserted in a mixture with graphene-oxide nanosheets. The authors proposed a switching mechanism based on the formation and dissolution of conductive filaments due to the migration of oxygen ions and their reabsorption from the mixture into an insulating layer appearing at the Al electrode/mixture interface. Most of articles [10], [19], [20] suggest that diffusion of ions from metal electrode into sulphur vacancies forms conductive filament. Regarding devices based on $MoS_2$ nanosheets, Tang et al. [21] explained the electrical switching by the migration of sulphur vacancies only at the edge of the nanosheets. Finally, Zhang et al. [22] justified the switching of the devices by a local phase change. None of these hypotheses has been formally confirmed. Moreover, several of them could coexist in the same device.

Memristor performances and our understanding of switching mechanisms are currently limited by the low yield of switching devices, lack of reproducibility and weak endurance after manufacturing [23]. Developing a 2D material integration process with CMOS compatibility remains challenging and still limits the performance of devices for memory as well as for other applications (eg. photoFET [24] or sensors [25]). Because of their unique structure, 2D materials are particularly sensitive to the standard integration processes of semiconductor industry [26]. The following points are particularly critical and have been highlighted in recent years: contamination with polymer residue [27], [28], how to manage Van der Waals interfaces [12], [29] Contact engineering on 2D TMDs [30], [31], [32].

In this study, we have developed and optimized a complete process for integrating 2D materials into a fully CMOS-compatible vertical MIM structure in a clean room and over large area. The preservation of the physico-chemical properties of $MoS_2$ is assessed along the whole fabrication process. Electrical measurements are performed to study the influences of the fabrication protocol, the top electrode metal and the $MoS_2$ grain size on devices performances. Eventually, switching mechanisms are discussed in terms of conduction through grain boundaries.

**2. Experimental methods**

*2.1 MoS$_2$*

Two types of $MoS_2$ were integrated in the devices. $MoS_2$-A sample corresponds to three monolayer of $MoS_2$ deposited by Atomic Layer deposition (ALD) onto 8'' Si/$SiO_2$ wafer with $Mo(NMe_2)_4$ and 1,2-ethanedithiol as precursors at around 100°C. An annealing treatment at 900°C under inert atmosphere is performed to improve the crystallinity. The detailed method is given in Ref [33]. $MoS_2$-B sample is provided by 2D semiconductor. The deposition was done by CVD onto a 2'' sapphire substrate. $MoS_2$-B sample corresponds to one monolayer [34].

*2.2 Characterization*

X-ray photoelectron spectrometry (XPS) measurements were performed with a PHI 5000 VersaProbe II photoelectron spectrometer at the Nanocharacterization Platform (PFNC) of the Minatec Campus in Grenoble (France). The samples were inserted in ultrahigh vacuum and excited by a 200 μm diameter monochromatic Al-Kα (hν = 1486.6 eV) X-ray light source. The source analyser angle is 45° and remains unchanged during the XPS data acquisition, where the photoelectron take-off angle (sample-analyser angle) also fixed at 45° does not vary either. The overall energy resolution (taking into account both the spectrometer and X-ray bandwidths) is 0.6 eV for Core Level spectra and we set the C 1s peak from the adventitious carbon to 284.8 eV for calibration of the binding energy scale. Data correction and peak deconvolution were performed with the CasaXPS software, and we used the sensitivity factors provided by the instrument to perform the relative quantification of the elements.

Atomic force microscopy (AFM) measurements were performed on a Dimension ICON (Bruker) in PeakForce Tapping mode using a "ScanAsyst Air" cantilever (nominal stiffness: 40 N/m; tip radius of curvature: 5 nm; and approach-withdrawal frequency: 2 kHz).

Raman spectroscopy spectra were obtained using a Renishaw inVia Reflex spectrometer with a 532 nm laser diode excitation source coupled with a 1800 grooves/mm grating and on 100× objective (NA = 0.9). The laser power is limited to < 0.1 mW (meausred at sample's surface) to advoid any laser induced heating effect. The spectra were recorded over a range of 100-1000 $cm^{-1}$. To determine width and peak positions, the curves were fitted using the same method as in Mignuzzi et al [35].

Cross-section specimen for transmission electron microscopy (TEM) were prepared using a Zeiss crossbeam 550 FIBSEM.

Scanning Transmission electron microscopy (STEM) analyses were performed using Cs-corrected FEI-Titan Themis microscope, operating at 200 kV and allowing high angle annular dark field (HAADF) imaging. 4-dimensional (4D) STEM dataset was acquired using a direct electron detector (Quantum Detector, Medipix-Merlin) and the reconstructed orientation map was obtained by the three-fold filtering as demonstrated in [36].

Electrical measurements were performed in ambient conditions under two tungsten tips using a Keysight B1500A Semiconductor Analyser with: a B1517A High Resolution Source Measurement Unit. Each measurement involves applying a voltage sweep between the two tips and measuring the current. To control the formation of the conductive filament, a compliance current was set during the tuning operation, the value of which was approximately $10^{-5}$ A.

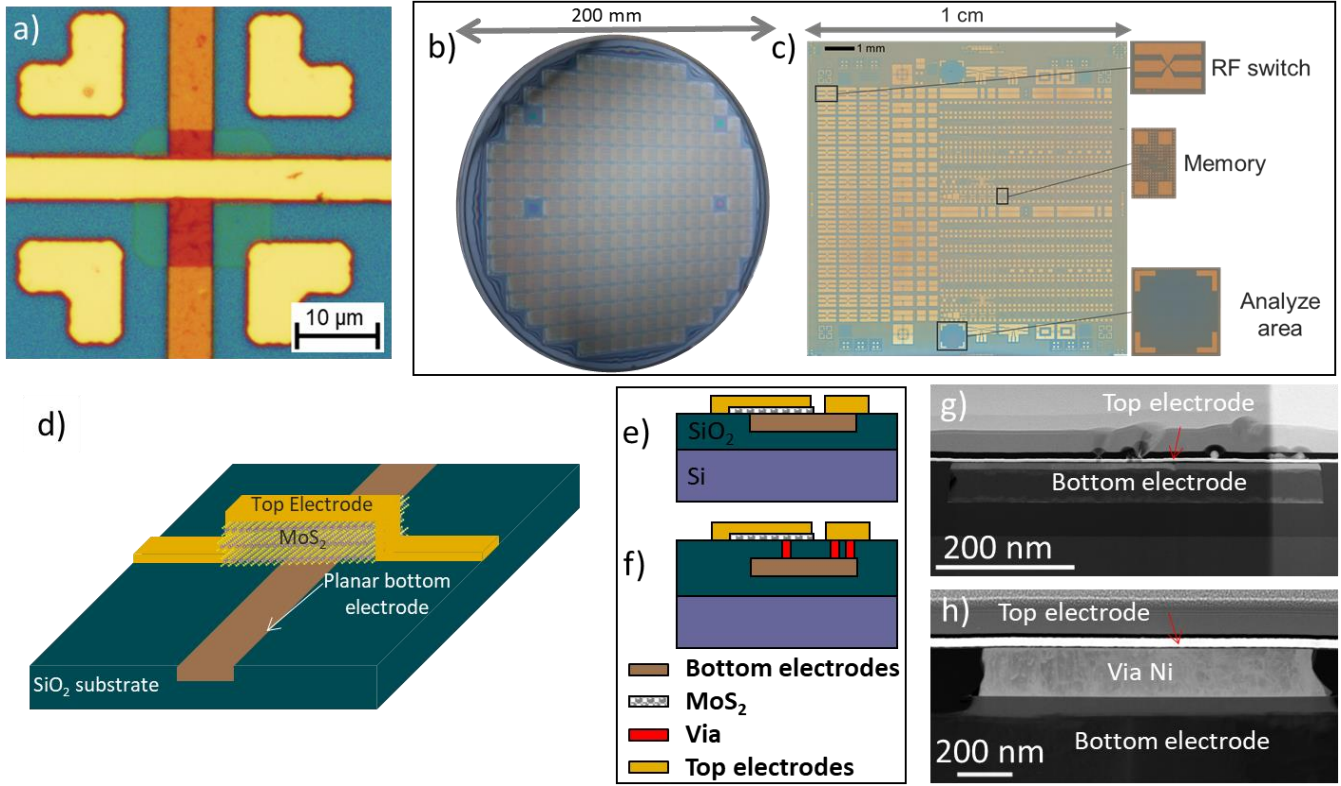

*Figure 1: Integration process: Image of the 200 mm wafer including the back electrodes (b) with about 250 dice, Optical image of one die with some detailed devices (c) memristor devices schematics (d), optical image (a) and cross-section representation of the 2 architectures of bottom electrode without (e) and with via (f). Cross-section TEM images of the architectures without (g) and with via (h).*

## 2.3 Device fabrication

The goal of the integration process is to develop vertical cross memory devices as illustrated in Figure 1-a and d. A first protocol called protocol I was used to manufacture the first devices. It was then optimized to protocol II for devices with improved performances.

### 2.3.1 Bottom electrodes

Planar bottom electrodes (obtained by chemical-mecanical polishing) are manufactured on 200 mm wafers in clean room using standard microelectronic processes. Using planar electrodes avoids the presence of defects related to the $MoS_2$ transfer onto 3D electrodes where step edges might break the film continuity. Each wafer contains 250 dice with various designs as presented in Figure 1-a. Several devices are patterned on each chip but only memories (cross devices) are used in this work (Figure 1-c, d). Two types of bottom electrodes are considered. In protocol I, they consist in the following stack: Ti(10 nm)/AlCu(440 nm)/Ti(10 nm) and TiN (100 nm) on top (Figure 1-e and g). In protocol II nickel via were integrated on top of the TiN electrode (Figure 1-f and h). The typical size of the active part of the cross-devices without and with via is about 25 µm² and 1 µm², respectively. By reducing the active surface area of the device with via we hope to improve the reproducibility of devices. In addition, it has been demonstrated [37] that reducing the size of the electrodes with via contributes to lowering the capacitance and improving the performance of RF switches, the targeted application. For the rest of the procedure, the 8'' wafer is cut into smaller dice of 9 cm². The following steps are summarized for each protocol in Table I.

*Table I : Detailed process integration steps for protocol I and protocol II.*

|  | **Potocol I** | **Protocol II** |
|---|---|---|
| $MoS_2$ transfer | Wet transfer (polymer assist) | Wet transfer (polymer metal assist) |
| $MoS_2$ patterning | Photolithography | Deep UV lithography |
|  | ICP | IBE |
| Top electrodes deposition and patterning | Photolithography | Deep UV lithography with double PMMA layer |
|  | EBPVD | EBPVD |
|  | Lift-off | Optimized lift-off |

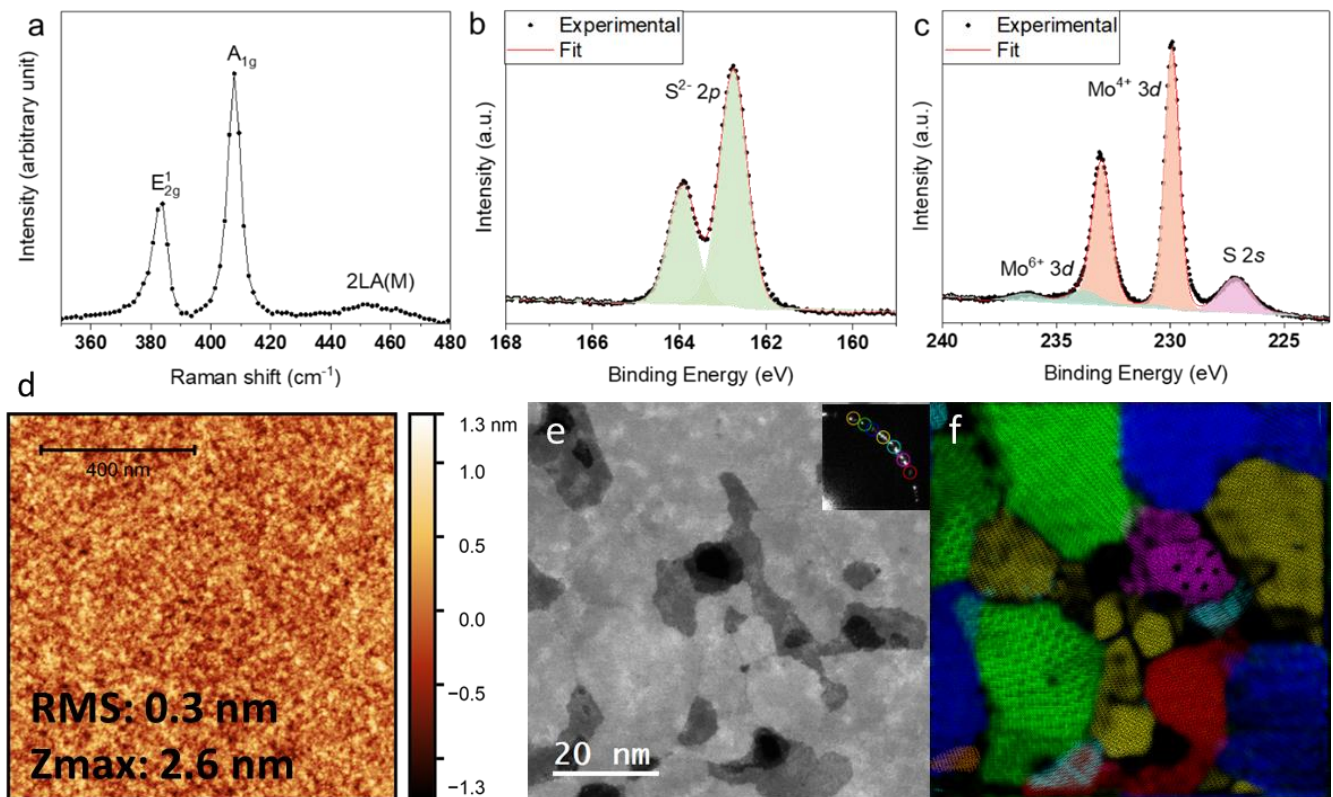

*Figure 2 : Physico-chemical characterization of $MoS_2$-A. Raman spectroscopy spectrum. (a) S 2p (b) and Mo 3d (c) XPS spectra. AFM image (d). Plane view STEM HAADF image of three monolayers $MoS_2$ with the corresponding fast Fourier transform (FFT) in the inset ((e). Dark field image reconstructed from selected spots in FFT corresponding to different crystal orientations(f). The attribution of false colors is indicated in the inset of (e).*

### 2.3.2 Transfer and patterning .of $MoS_2$

In protocol I, the $MoS_2$ transfer is performed using the classical wet transfer with PMMA as support layer as described in [38]. A solution of liquid polymethyl methacrylate (PMMA) is spin-coated on the surface of the $MoS_2/SiO_2/Si$ sample and baked at 180°C for 5 min to remove the solvent. The $MoS_2/SiO_2/Si$ sample is then dipped into water to detach the $MoS_2$/PMMA stack from the Si substrate. The stack is fished with dice of few square centimeters. The sample is dried in air and the PMMA removed by acetone. In protocol II, a thin Au-layer is deposited by evaporation on top of $MoS_2$ and the same process as $MoS_2$ only is applied. In protocol I, $MoS_2$ is patterned by a standard photolithography step using AZ1512HS as photoresist and AZ developer followed by an inductively coupled plasma (ICP) with $CF_4$. In protocol II a deep UV lithography is used with PMMA as photoresist., followed by an ion beam etching (IBE) step.

### 2.3.3 Top electrodes

Finally, top electrodes are patterned by lift off. For protocol I, AZ1512HS photoresist is spin coated, exposed and developed, then metal is deposited by electron beam physical vapor deposition (EBPVD) and patterned by photoresist removing in acetone. Three types of electrodes are used for the devices without via: Ni-10nm, Ti-10 nm/Pt-90 nm and Cr-2 nm/Au-98 nm. Ti and Cr are used as adhesive layers on $SiO_2$. For devices with via, in protocol II, only 100 nm of Au are used. The resist and lift-off solvent used for this process are PMMA and acetone respectively. The effect of an annealing at 300°C under an inert atmosphere of nitrogen and during 2 hours is studied to remove residual organic contamination and improve interfaces quality.

## 3 Results and discussion

### 3.1 $MoS_2$ characterization

#### 3.1.1 $MoS_2$-A

Raman spectroscopy was used to check the crystal quality, homogeneity, and number of $MoS_2$ layers in an area of 7x7 µm² comparable to the size of the active layer in cross devices. First, 49 scans were recorded to control the sample homogeneity. All the spectra superimpose (see Supplementary Figure S1). Figure 2-a exhibits the two MoS$_2$ characteristic vibration modes $E^1_{2g}$ and $A_{1g}$ at 383.2 and 407.7 cm$^{-1}$ respectively and also the 2LA(M) mode at 452.0 cm$^{-1}$ [39].The Raman peaks are very narrow, typically 5.4 and 5.3 cm$^{-1}$ for modes $E^1_{2g}$ and $A_{1g}$ respectively. It demonstrates the good crystalline quality of the film [35]. According to Li et al. [40], the gap between the two main peaks changes as a function of the number of layer. A gap of 24.5 cm$^{-1}$ was observed which corresponds to 4 monolayers (ML).

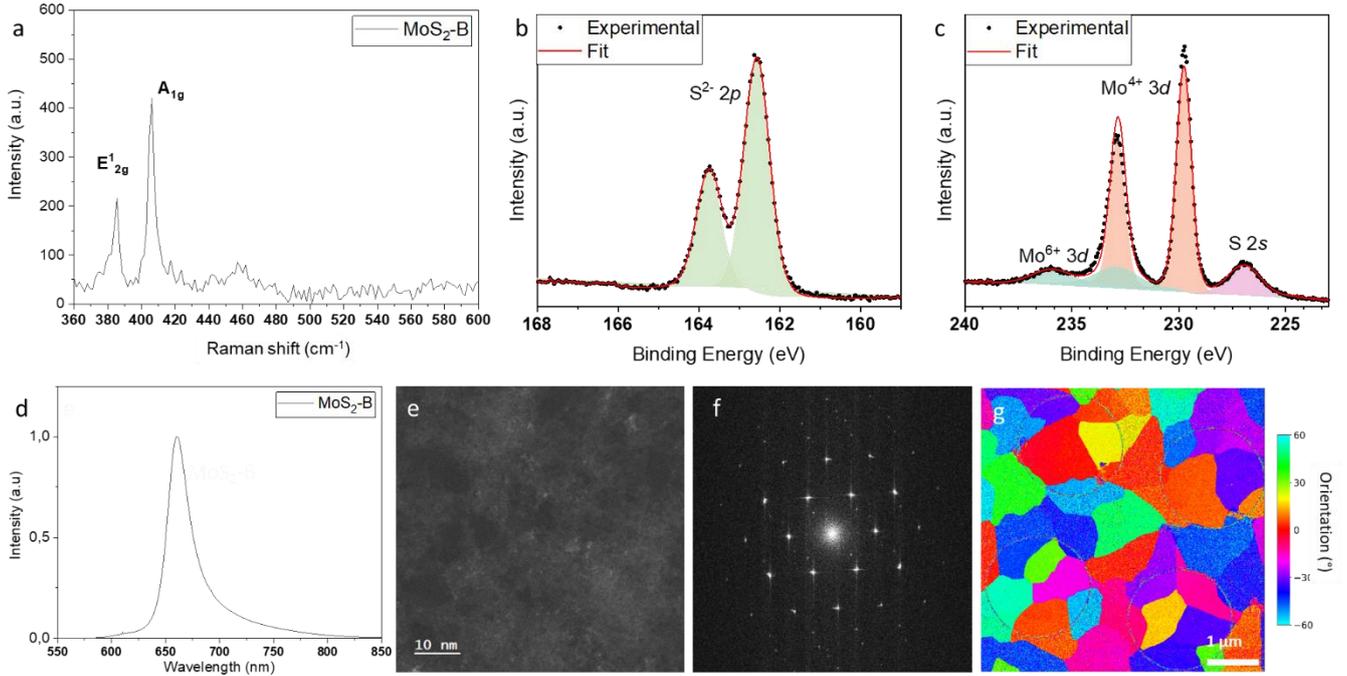

*Figure 3: Physico-chemical characterization of MoS$_2$-B. Raman spectroscopy spectrum (a). S 2p (b) and Mo 3d (c) XPS spectra. photoluminescence spectrum (d), Plane view STEM HAADF of monolayer of MoS$_2$ (e), the corresponding fast Fourier transform image (f) and large scale 4D-STEM orientation map visualizing grain distribution (g.).*

Rutherford Backscattering spectroscopy (RBS) and X-ray photoelectron spectrometry (XPS) were performed to study the chemical composition of the material. RBS (Supplementary Figure S2) reveals a slight deficiency in sulphur since the S/Mo ratio was determined as 1.76 ± 0.15. This sulphur deficiency is confirmed by XPS quantifying an atomic ratio between Mo $3d^{4+}$ and S $2p$ orbitals of 1.8. (Supplementary Table 1). Figure 2.-b and c show the high resolution XPS spectra of Mo $3d$ and S $2p$, respectively including deconvolution curves. The most intense peaks located at 229.9 and 233.0 eV are attributed to the doublet of Mo$^{4+}$ $3d_{5/2}$ and Mo$^{4+}$ $3d_{3/2}$ corresponding to Mo-S bonding, in agreement with the presence of 2H-MoS$_2$ [41]. Another doublet is visible on the spectrum at the following positions: 234 and 236.6 eV corresponding to the doublet of Mo$^{6+}$ $3d_{5/2}$ and Mo$^{6+}$ $3d_{3/2}$ [42]. These peaks indicate the presence of Mo-O bonds as a consequence of MoS$_2$ weak oxidation due to air exposure[43]. In the S $2p$ spectrum, two peaks are observed at 162.7 and 163.9 eV corresponding to the doublet of S$^{2-}$ $2p_{3/2}$ and S$^{2-}$ $2p_{1/2}$ respectively. These peaks are also attributed to Mo-S bonding [44].

Atomic force microscopy (AFM) was used to study the film morphology. A 1x1 µm² image is shown in Figure 2-d. The measured root-mean-square (RMS) roughness is as low as 0.3 nm for a surface of 1x1 µm² and the maximum peak-valley value is 2.6 nm. It shows the very low roughness of MoS$_2$-A.

Finally, STEM was used to probe the atomic structure of MoS$_2$-A. The dark field image shown in Figure 2-f demonstrate the polycrystalline nature of the film with an average domain size of 10 nm and random plane orientation of MoS$_2$ crystals.

### 3.1.2 MoS$_2$-B

MoS$_2$-B was characterized using the same methods. Figure 3-a exhibits the two MoS$_2$ characteristic vibration modes $E^1_{2g}$ and $A_{1g}$ at 385.3 and 405.0 cm$^{-1}$ respectively and also the 2LA(M) mode at 440.5 cm$^{-1}$ The Raman peaks are very narrow, typically 5,2 and 4.6 cm$^{-1}$ for $E^1_{2g}$ and $A_{1g}$ respectively. It demonstrates the good crystallinity of the film [35]. The gap between the two main peaks is 19.7 cm$^{-1}$, according to Li et al. [40], it corresponds to 1 monolayer.

Figure 3-b and c show the high resolution XPS spectra of Mo $3d$ and S $2p$ including the deconvolution curves. By XPS we find an atomic S/Mo ratio of 2 based on Mo $3d$ and S $2p$ signals. The binding energy gap between Mo $3d_{5/2}$ and S $2p_{3/2}$ of MoS$_2$-B as 67.3 eV which confirms the good

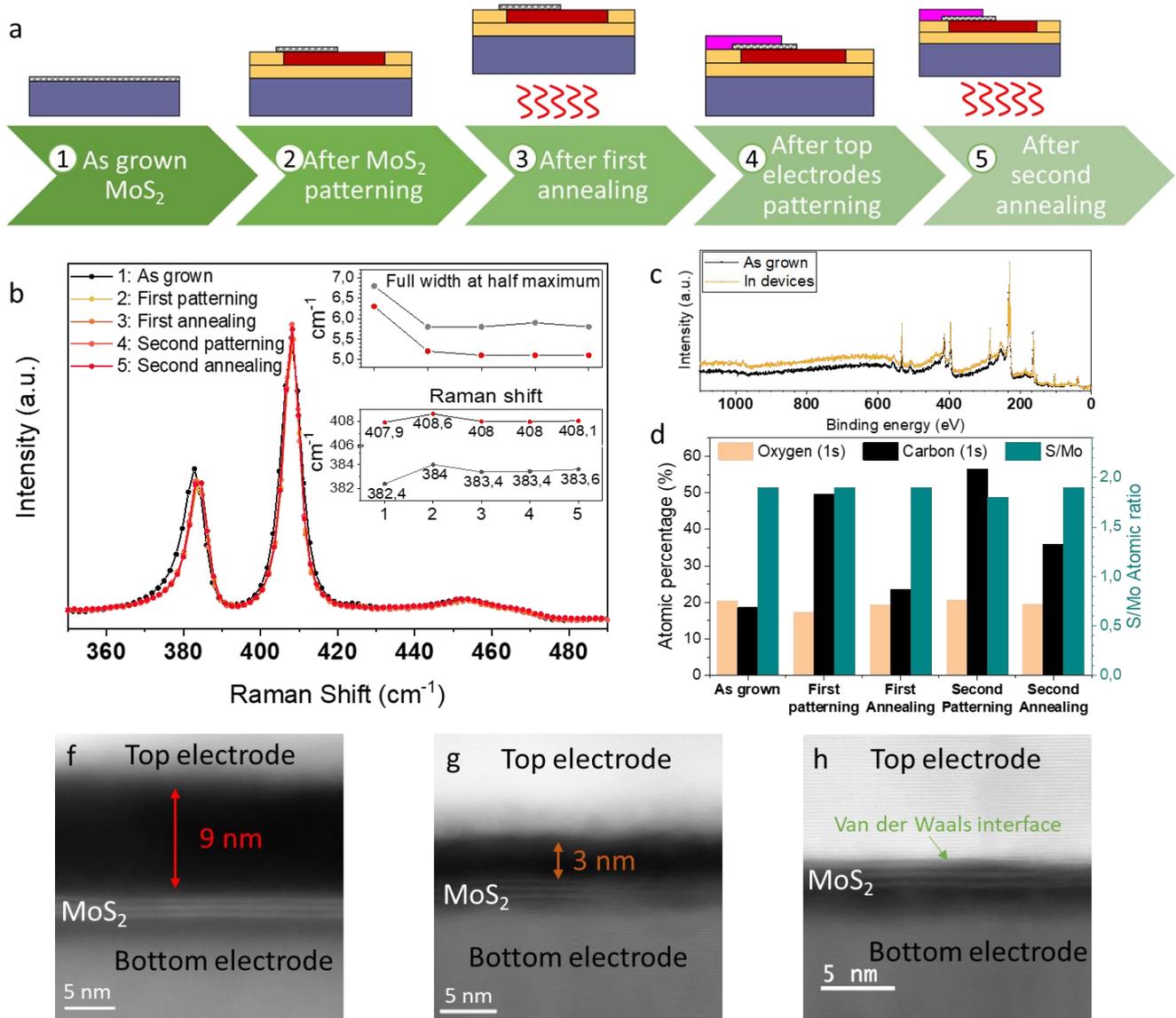

*Figure 4: Study of integration process impact: (a): Schematic of the different process steps. (b): Raman spectra during the process, inset evolution of $E^1_{2g}$, in grey, and $A_{1g}$, in red, peak position and full width at half maximum (FWHM). (c): XPS survey spectra of MoS₂ as grown (in black) compared with MoS₂ incorporated in devices (in yellow) and (d): evolution of oxygen, carbon contents and S/Mo⁴⁺ ratio during the process. (f-h): Cross-section TEM image of MoS₂ integrated in the devices at the end of the process without annealing (f) with annealing (g) with transfer with Au layer (h).*

stoichiometry of MoS$_2$. Figure 3-d shows photoluminescence spectra [30]. The intensity and low FWHM of the lines highlight a typical well-crystallized MoS$_2$ monolayer. These observations are confirmed by the TEM images. Figure 3-e and f show a continuous and monocrystalline layer on small area (around 100x100 nm²). Orientation map obtained in large area (5x5 µm²) demonstrates a continuous layer with domain size of ~1 µm² (Figure 3-g).

*3.2 Impact of the fabrication process*

The impact of the integration process is studied on devices made with protocol I. The analyze area, visible on Figure 1-b, was characterized by Raman spectroscopy and XPS at each step described in Figure 4-a and by STEM analysis at the end of the process. The insets in Figure 4-b report the peak position and FWHM at each process step. Raman spectra exhibit almost no variation of the MoS$_2$ vibration peak positions and FWHM except after transfer and the first patterning. We attribute it to stress release. Actually, MoS$_2$ is subjected to an important thermal stress during the whole preparation process and the transfer onto another substrate allows to release the strain [45]. The Raman analysis highlights that the MoS$_2$ structure is preserved during the device preparation without any observable damage. The XPS analysis (Figure 4-c and 4-d) shows that the S/Mo4+ ratio and oxygen content remain constant during the process. This proves that the MoS$_2$ stoichiometry is preserved, and the material is not oxidized during the process. On the other hand, the carbon content is increased after each lithography step and reduced by annealing. Indeed, the lithography and transfer resins bring carbon contamination onto MoS$_2$. This layer cannot be removed by usual methods as dipping in a basic solution or plasma treatment due to the sensitivity of MoS$_2$ to these processes. This observation is confirmed by TEM analysis without (Figure 4-f) and with (Figure 4-g) annealing. A carbon contamination layer of 9 nm is observed while with annealing it is reduced to 3 nm. In protocol II, to avoid carbon contamination at the interface with the top electrode, MoS$_2$ was transferred using a thin layer of gold on top, which also ensure a good electrical contact between MoS$_2$ and the top electrode. The cross section shown in Figure 4-h has been performed on a stack with Ni-via/MoS$_2$-A/Au. The image shows the perfect interface between MoS$_2$ and the electrodes with no carbon contamination. This figure also exhibits the 3 continuous layers of MoS$_2$ and then that the MoS$_2$ structure has been preserved during Au deposition. The MoS$_2$ preservation has also been confirmed by Raman spectroscopy. Other works demonstrated that evaporation is a suitable process to deposit a film on top of 2D materials [46], [47].

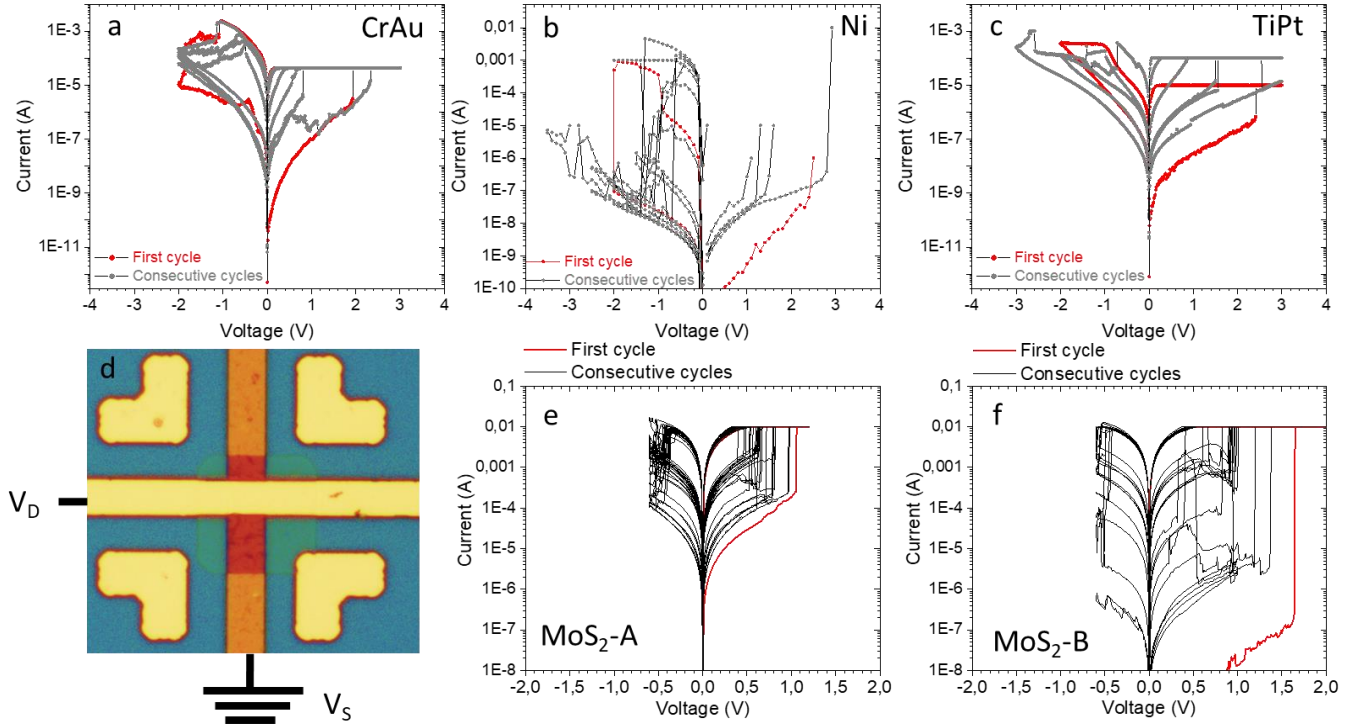

*Figure 5: Electrical results: I(V) characteristic of devices made with protocol I with various metals tested as top electrodes: CrAu (a), Ni (b), TiPt (c). Geometry of the measurement set up (d). I(V) characteristics of devices made with protocol II with MoS$_2$-A (e) and B (f)).*

In summary, protocol I preserves the MoS$_2$ crystal quality and stoichiometry but introduces carbon contamination at the interface. Such a carbon contamination has often been reported in the literature [27], [28], [48] but its full removal has never been achieved. Protocol II allows to avoid this contamination, yielding clean and sharp interfaces with the transferred MoS$_2$ layer.

*Electrical results*

Three metals were tested as top electrodes on devices using protocol I: gold with a chromium adhesion layer, nickel and titanium layer with platinum on top for high conductivity. For this first fabrication process, only 75% of the devices were correctly patterned. In addition, switching loops were observed only for few devices probably due to carbon contamination at the MoS$_2$/top electrode interface. For each metal, switching loops measured on a single device are shown in Figure 5a-c. A simple electrical set up was used to apply voltage sweeps and measure the current (Figure 5-d). Switching loops are observed for all the devices with all three metals used as top electrodes. All the devices were initially in a resistive OFF state and successfully switched to a reversible ON state. Switching from OFF to ON is called Set, while switching from ON to OFF is called Reset. The I(V) characteristics exhibit a lack of reproducibility between devices and loops. Nevertheless, a metal-dependent behaviour, consistent with the metal ion diffusion switching mechanism [10], [49], is observed. Indeed, the current ratio between ON and OFF states ($I_{ON}/I_{OFF}$) is about $10^2$ for CrAu and TiPt electrodes whereas it systematically reaches $10^5$ for nickel electrodes. The best endurance is observed for nickel and CrAu devices for which 7 consecutive loops were observed before the devices got locked in the ON state. The same measurements were carried out on devices made with Protocol II. For these samples, 100% of the devices were correctly patterned and switching loops were observed on a larger number of devices allowing to establish a yield of "switching devices". This yield was calculated by counting the number of devices capable of performing two full switching loops as the first one could be due to a shortcut introduced during the transition from OFF to ON and a device destruction for the transition from ON to OFF. For MoS$_2$-A and B, the yield of switching devices is around 40%. Typical switching loops are shown in Figure 5-e and f for MoS$_2$-A and MoS$_2$-B respectively. A better reproducibility is observed between devices and loops compared to devices made with protocol I. This is particularly true for MoS$_2$-A which shows a better reproducibility but lower current ratio than MoS$_2$-B. For both materials, the endurance is improved to 20 cycles. The higher rate of switching devices, the better reproducibility between devices and the improved cyclability demonstrate the importance of the integration process quality.

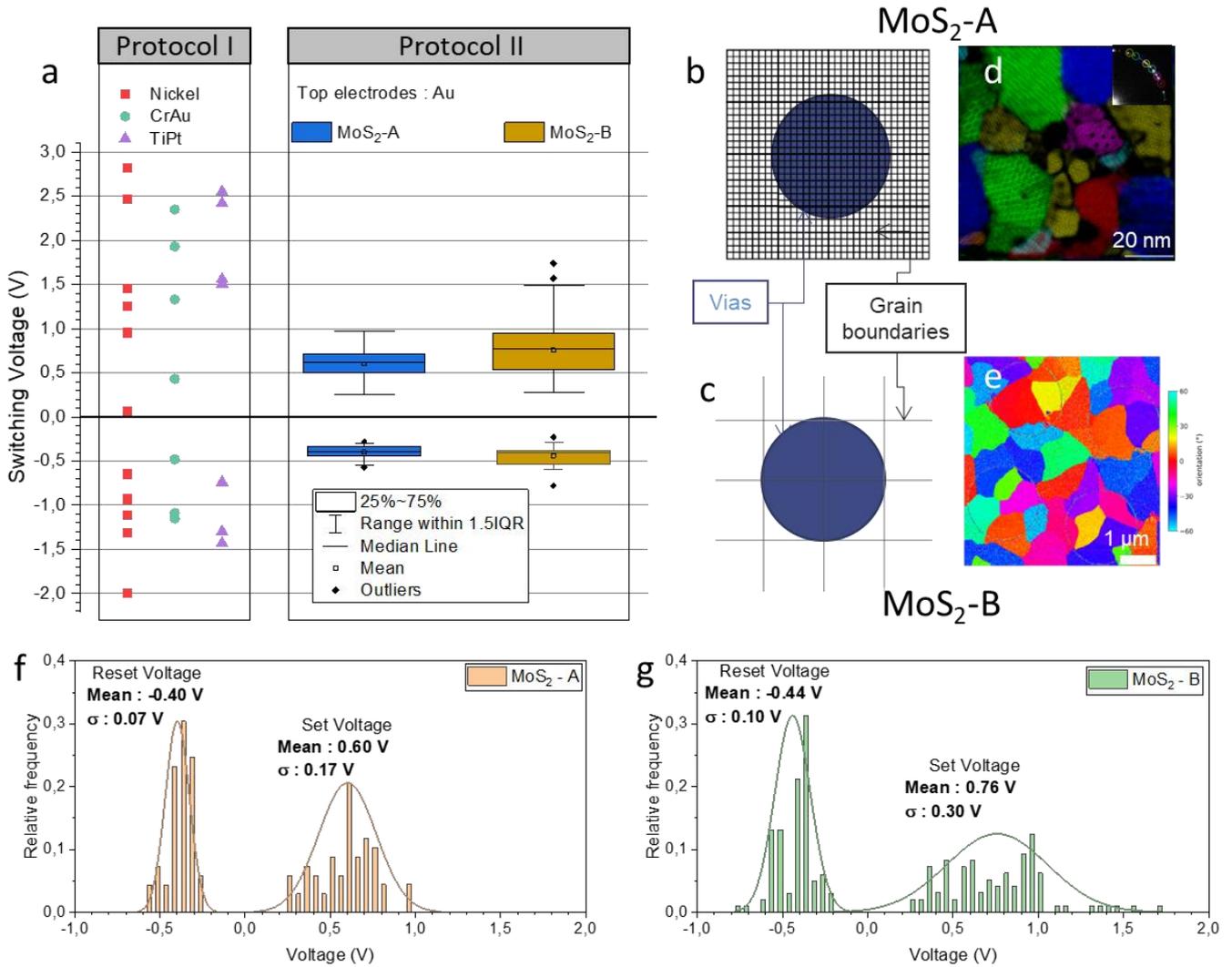

*Figure 6 : (a): Statistical comparison of switching voltages for devices made with protocol I (plotted with dots) and with protocol II (plotted with boxes). These statistics were established on 8 devices and 68 data for MoS$_2$-A and 12 devices with 98 data for MoS$_2$-B. Schematics of the grain boundaries density on via for MoS$_2$-A (b) and B (c). Grain size illustrated by fast Fourier transform in the inset showing MoS$_2$ crystalline diffraction pattern and: corresponding spatial distribution of grains in false colours for MoS$_2$-A (grain average diameter of ~10 nm)(d) and 4D-STEM mapping for MoS$_2$-B (grain average diameter of ~1 μm) (e). Histograms and normal distributions of switching voltages for MoS$_2$-A (f) and B (g)*

By controlling the integration process, we achieve good reproducibility from one device to another. As a consequence, a full set of data is collected on several devices providing enough statistics to study in detail the switching mechanism.

*Switching mechanism study*

To better understand the behaviour of the devices and the switching mechanism, the switching voltages were measured and statistically analysed for each sample. They are shown in Figure 6-a. For the samples made with protocol I, the lack of data and reproducibility made impossible to draw up a clear statistics and dots were plotted for each measurement. The distribution is broad and stochastic unlike the switching voltages measured for samples made with protocol II. For the latter, the data are represented by box plots for set and reset voltages that extend over thinner areas. This confirms that optimizing the process improves the performances of the devices in terms of reproducibility and endurance. It also allows the switching voltage to be lowered, thereby reducing devices power consumption. Between Protocol I and II, 3 integration parameters were modified: device size (from 25 µm$^2$ to less than 1 µm$^2$ with via in protocol II), transfer method and lithography procedure. So, it is not possible to determine the role of each parameter on devices performances.

These statistical results give the opportunity to study the switching mechanism by comparing the experimental data for MoS$_2$-A and B. This study is based on the work by Tang et al.[21] in which they demonstrated that the preferred location for switching is at the edge of MoS$_2$ nanosheets where crystalline defects are concentrated. In our samples, the crystalline defects are concentrated at grain boundaries. The MoS$_2$-A grain size is of order of 10 nm while the one of MoS$_2$-B it is about 1 µm. Thus, in MoS$_2$-A, the density of grain boundaries is significantly higher than in MoS$_2$-B. This is schematized and illustrated in Figures 6-b to 6-e. As there are significantly more crystalline defects in MoS$_2$-A, this material should switch more easily than MoS$_2$-B. This would result in a lower energy required to form a conductive filament and a lower switching voltage with a narrow distribution. For devices containing MoS$_2$-B, only few grain boundaries are present on the active area (Figure 6-c) i.e. there are few sites available to for the formation of conductive filaments. As a result, the switching voltage should be higher and with a larger distribution compared with MoS$_2$-A. The histograms and normal distributions of Set and Reset voltages for the two materials shown in Figure 6-f and 6-g confirm this assumption since the average set voltage is 0.60 V for MoS$_2$-A vs. 0.76 V for MoS$_2$-B with standard deviation of 0.17 V and 0.30V respectively. Similarly, the average reset voltages are – 0.4 V vs. – 0.44 V for MoS$_2$-A and B respectively with respective standard deviations of 0.07 V and 0.10 V. These results are consistent with the mechanisms proposed by Tang et al. In this study they demonstrated that despite the random nature of filament formation, this phenomenon is facilitated by the presence of defects in the MoS$_2$ structure. In our case, defects are mainly located at grain boundaries and the material with higher grain boundaries density switches more easily than the one with large grains. These defects could be sulphur vacancies, since XPS quantification measures less sulphur in MoS$_2$-A and since several simulation studies demonstrate the importance of theses vacancies for the switching mechanism [50], [51]. From this result, we can conclude that defects would be beneficial for vertical memory devices as they allow to obtain lower and more reproducible switching voltages. Then, MoS$_2$ with crystalline defects could be preferred for memristor manufacturing or pristine MoS$_2$ could be artificially damaged to enhance memristor performances [23], [52], [53].

## 5. Conclusion

In summary, we have developed a large-scale cleanroom-compatible process for the integration of ultrathin MoS$_2$ layers in vertical MIM devices. We demonstrate the preservation of MoS$_2$ crystalline integrity, stoichiometry, and the achievement of high-quality Van der Waals interfaces by optimising the process.

Thanks to these optimisations, the yield of switching devices as well as reproducibility and performances are improved. More specifically, an ON/OFF current ratio of 10$^5$ is obtained, which is promising for the manufacture of RF switches in the future. By reducing the stochastic behaviour of the devices, we could confirm hypothesis regarding the switching mechanism. Statistics on the switching voltage and distribution of MoS$_2$ devices with two different grain sizes demonstrate that grain boundaries are the preferred location for the conductive filament formation. Therefore, the quality of MoS$_2$ layers, particularly the grain size, is a key factor to adjust the performances of MoS$_2$-based memories.

**Acknowledgements**

This work was supported by the French Public Authorities within the framework of the Nano2022 (IPCEI Microelectronics) project. The authors thank the "Plateforme Technologique Amont" of Grenoble, with the financial support of the CNRS Renatech network. Finally, a part of this work, carried out on the NanoCharacterization Platform (PFNC), was supported by the Recherches Technologiques de Base" Program of the French Ministry of Research. The 4D-STEM analysis has been supported by the funding from the European Research Council under the European Union's H2020 Research and Innovation programme via the e-See project (Grant No. 758385).

**Data availability statement**

All data that support the findings of this study are included within the article (and any supplementary files).

**Data access statement**

All relevant data are within the paper.

**Conflict of interest**

The author does not have any conflict of interest with the work reported in this manuscript.